\documentclass[twocolumn, showpacs,preprintnumbers,amsmath,amssymb]{revtex4}
\usepackage{amssymb}

\usepackage{graphicx}
\usepackage{dcolumn}
\usepackage{bm}
\usepackage{multirow}
\usepackage{booktabs}
\usepackage{CJK}
\usepackage{xcolor}

\begin{document}
\preprint{APS/123-QED}

\title{Optical shielding of destructive chemical reactions between ultracold ground-state NaRb molecules}
\begin{CJK*}{GBK}{song}\end{CJK*}

\author{\begin{CJK*}{GBK}{song}T. Xie$^1$\end{CJK*}}
\author{\begin{CJK*}{GBK}{song}M. Lepers$^2$\end{CJK*}}%
\author{\begin{CJK*}{GBK}{song}R. Vexiau$^1$\end{CJK*}}%
\author{\begin{CJK*}{GBK}{song}A. Orb\'{a}n$^3$\end{CJK*}}
\author{\begin{CJK*}{GBK}{song}O. Dulieu$^1$\end{CJK*}}
\author{\begin{CJK*}{GBK}{song}N. Bouloufa-Maafa$^1$\end{CJK*}}

\affiliation{$^1$Universit\'e Paris-Saclay, CNRS, Laboratoire Aim\'{e} Cotton, 91405 Orsay, France}%
\affiliation{$^2$Laboratoire Interdisciplinaire Carnot de Bourgogne, CNRS, Universit\'{e} de Bourgogne Franche-Comt\'{e}, 21078 Dijon, France}%
\affiliation{$^3$Institute for Nuclear Research, Hungarian Academy of Sciences(ATOMKI), H-4001 Debrecen, Pf. 51, Hungary}%

\date{\today}

\begin{abstract}

We propose a method to suppress the chemical reactions between ultracold bosonic ground-state $^{23}$Na$^{87}$Rb molecules based on optical shielding. By applying a laser with a frequency blue-detuned from the transition between the lowest rovibrational level of the electronic ground state $X^1\Sigma^+ (v_X=0, j_X=0)$, and the long-lived excited level $b^3\Pi_0 (v_b=0, j_b=1)$, the long-range dipole-dipole interaction between the colliding molecules can be engineered, leading to a dramatic suppression of reactive and photoinduced inelastic collisions, for both linear and circular laser polarizations. We demonstrate that the spontaneous emission from $b^3\Pi_0 (v_b=0, j_b=1)$ does not deteriorate the shielding process. This opens the possibility for a strong increase of the lifetime of cold molecule traps, and for an efficient evaporative cooling. We also anticipate that the proposed mechanism is valid for alkali-metal diatomics with sufficiently large dipole-dipole interactions.
\end{abstract}


\maketitle

Ultracold quantum gases composed of particles interacting at large distances in an anisotropic manner represent promising platforms for studying many-body physics and strongly-correlated systems \cite{baranov2008,bloch2008,lahaye2009} for high-impact applications like quantum simulation \cite{bloch2012} and quantum computation \cite{leshouches2009}. Among the possible candidates feature ultracold polar molecules, as their large permanent electric dipole moment induced by an electric field can generate intense long-range anisotropic interactions \cite{doyle2004,carr2009,dulieu2009,quemener2012,moses2017,bohn2017}. However, such studies with ultracold molecules in their absolute ground state are hindered when ultracold chemical reactions occur at short distances, like with the $^{40}$K$^{87}$Rb species \cite{ospelkaus2010a,ni2010}. Even for the non-reactive bosonic $^{87}$RbCs \cite{takekoshi2014,molony2014,gregory2019} and $^{23}$Na$^{87}$Rb \cite{guo2016}, or fermionic $^{23}$Na$^{40}$K \cite{park2015,seesselberg2018} species, which were thought to be immune to inelastic losses \cite{zuchowski2010} in their absolute ground state, limited lifetimes of the ultracold molecular samples were recorded. The origin of this loss mechanism is tedious to identify as the final products are not easily detected \cite{guo2016,guo2018}. However, a first success in this direction has been recently obtained with the reactive species $^{40}$K$^{87}$Rb \cite{hu2019,liu2020}.

In their theoretical study, Christianen \emph{et al}.~\cite{christianen2019} suggested that the observed losses in NaK + NaK collision come from the electronic excitation caused by trapping lasers even though they are far-detuned. This hypothesis has been confirmed when Gregory \textit{et al.} observed that the dominant loss process in ultracold $^{87}$RbCs collisions arises from the fast optical excitation of long-lived two-body complexes \cite{gregory2020}. They have shown that the losses may be significantly suppressed by square-wave modulation of the trap intensity. However this is still not enough to prepare long-lived ultracold molecular samples in the absolute ground state.

Instead, one can use electromagnetic (em) field without any additional field to control the long-range interactions between two ground-state bialkali molecules AB in their lowest vibrational level $v=0$. A microwave (mw) field is tuned on the transition between the two rotational sublevels $j=0$ and $j=1$. This engineers repulsive long-range interactions between AB molecules, due to the coupling between the entrance scattering channel AB($j=0$)+AB($j=0$) and the neighboring one AB($j=0$)+AB($j=1$) \cite{karman2018,lassabliere2018,avdeenkov2015,karman2020}. It was found however that a significant shielding can only be achieved with circularly-polarized mw-field, which is tedious to implement experimentally \cite{napolitano1997}.

An em field in the optical frequency range couples electronic states of the colliding particles. The proof-of-principle of such an optical shielding (OS) has been demonstrated for inelastic collisions between ultracold alkali-metal atoms, by coupling the $^2S$ ground state to the lowest $^2P$ excited state \cite{bali1994,marcassa1994,suominen1996a,hoffmann1996,zilio1996,muniz1997,weiner1999}. The circularly-polarized light was shown to work better than linear-polarized one, but the shielding efficiency was hindered by spontaneous emission from the $^2P$ level \cite{suominen1995,suominen1996b,zilio1996,napolitano1997}.

\begin{figure}[!t]\centering
\resizebox{0.45\textwidth}{!}{%
  \includegraphics{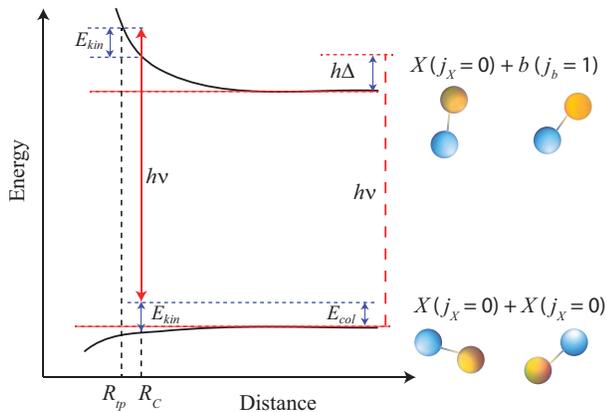}
}
    \caption{Classical picture of the optical shielding of ultracold molecular reactive collisions. Two molecules in the lowest rovibrational level of their electronic ground state $X(j_X=0)$ ($v_X=0$ is omitted for convenience) collide with energy $E_{col}$. At the Condon point $R_C$ (conserving the local kinetic energy $E_{kin}$ in this picture) the pair absorbs a photon of energy $E=h\nu$ to reach a repulsive long-range potential energy curve of a ($X(j_X=0)$) molecule interacting with an excited ($b(j_b=1)$) molecule ($v_b=0$ is omitted for convenience). The optical frequency $\nu=E/h$ is blue-detuned by $\Delta>0$ from the $X(j_X=0)$-$b(j_b=1)$ transition frequency $\nu_{Xb}$, so that $E=h\nu=h\nu_{Xb}+h\Delta$. The position $R_{tp}$ associated to the local kinetic energy $E_{kin}$ at $R_C$ is the classical turning point for the relative motion in the upper curve. The molecules are thus prevented from reaching the reactive zone. Stimulated emission in the excited complex also occurs if the field is strong enough (thus the double-arrowed red line).}
    \label{fig:scheme}
\end{figure}

Here, we demonstrate the OS efficiency to suppress reactive collisions between ultracold $^{23}$Na$^{87}$Rb molecules in the lowest rovibrational level ($v_X=0, j_X=0$) of the $X^1\Sigma^+$ ground state (noted $j_X=0$) in free space. The principle, illustrated in Fig.\ref{fig:scheme}, involves a laser with frequency blue-detuned from the transition between the $j_X=0$ level and the ($v_b=0, j_b=1$) rovibrational level of the excited $b^3\Pi_0$ state (noted $j_b=1$), with energy $E/hc=11306.5$~cm$^{-1}$ (or 884.447~nm) \cite{docenko2007}. This spin-forbidden transition, facilitated by the spin-orbit coupling between the $b^3\Pi_0$ and the $A^1\Sigma^+_0$ excited states, ensures that the spontaneous emission does not hinder the OS efficiency. We calculated the bimolecular long-range potential-energy curves (PECs) from on our previous developments \cite{lepers2013,vexiau2015,lepers2018,li2019} safely neglecting the hyperfine structure of the rotational levels (see Supplementary Material (SM) \cite{sm}). We found that a PEC correlated to the $^{23}$Na$^{87}$Rb($j_X=0$)+$^{23}$Na$^{87}$Rb($j_b=1$) asymptote possesses a high potential barrier caused by dipole-dipole interaction (DDI). In contrast with the microwave case, OS is efficient for both circularly ($\sigma^\pm$) and linearly ($\pi$) polarized light. Its efficiency increases monotonously with the strength of the $X$-$b$ optical coupling, and is almost temperature-independent. Note that OS is not perturbed if the molecules are held in a conventional optical dipole trap with a wavelength far off any resonance reachable from the $j_X=0$ level, like 1064~nm \cite{sm}.

We solve the time-independent Schr\"odinger equation describing the two-body collision between ground-state $^{23}$Na$^{87}$Rb molecules in the presence of an OS laser field, and no other field. The Hamiltonian for the bimolecular complex with reduced mass $\mu$ is written as
\begin{equation}
    \begin{aligned}
        \hat{H} = -\frac{\hbar^{2}}{2\mu R^{2}}\frac{\partial}{\partial R}\left(R^2 \frac{\partial}{\partial R}\right)+\frac{\bf{L}^{2}}{2\mu R^{2}}+\hat{H}_{\textrm{rot}}\\+\hat{V}(R)+\hat{H}_{f}+\hat{H}_{I},
    \end{aligned}
\label{eq:hamilt}
\end{equation}
where $\bf{L}$ is the angular momentum of the mutual rotation of the molecules with the quantum number $\ell$ (namely, the partial wave), $\hat{H}_{\textrm{rot}}$ is the sum of the rotational Hamiltonian for each molecule. The operator $\hat{V}(R)$ holds for the long-range interaction between the molecules, including the DDI scaling as $R^{-3}$ and the van der Waals (vdW) interaction scaling as $R^{-6}$ \cite{lepers2013,vexiau2015,lepers2018,li2019} (see Supplementary Material). The laser field Hamiltonian is $\hat{H}_f$, and $\hat{H}_I$ denotes the interaction between the complex and the laser field, depending on the Rabi frequency $\Omega$ and the detuning $\Delta$ with respect to the $(j_X=0)$-$(j_b=1)$ transition energy.

We introduce the symmetrized fully-coupled basis functions in the laboratory frame  $|e_1,j_1,p_1,e_2,j_2,p_2,j_{12},\ell,J,M\rangle$ to characterize the scattering channels  \cite{li2019}: for $i=1,2$, the quantum numbers $e_{i}=X,b$, $j_{i}$ and $p_{i}=\pm 1$ respectively describe the electronic state, the total angular momentum $\bf{j_{i}}$ and the parity of each molecule.  If $e_i=X$, then $p_i=(-1)^{j_i}$, while if $e_i=b$, both parities $p_i=\pm 1$ exist for a given $j_i$. The angular momenta $\bf{j_1}$ and $\bf{j_2}$ are coupled to give $\bf{j_{12}}$, itself coupled with $\bf{L}$ to give the total angular momentum of the complex $\bf{J}$ and its projection $J_z$ on the laboratory-frame $z$ axis (with quantum number  $M$).

We use the dressed-state model \cite{cohen-tannoudji1998} to account for photon absorption and stimulated emission, so we introduce the basis vector $|n\rangle$ associated to the number $n$ of photons of the OS field. It is sufficient to restrict the calculation within the Floquet block $|n=0\rangle$ for $X+X$, and $|n=-1\rangle$ for $X+b$ \cite{sm}. For bosonic molecules, the dressed symmetrized basis functions are related to the unsymmetrized ones by
\begin{equation}
    \begin{aligned}
     & |e_1,j_1,p_1,e_2,j_2,p_2,j_{12},\ell,J,M\rangle |n\rangle \\
     & = \frac{1}{\sqrt{2(1+\delta_{e_1e_2}\delta_{j_1j_2}
       \delta_{p_1p_2})}} \\
     & \times ( |e_1(j_1,p_1),e_2(j_2,p_2),j_{12},\ell,J,M\rangle
       |n\rangle \\
     & \quad + \varepsilon
       |e_2(j_2,p_2),e_1(j_1,p_1),j_{12},\ell,J,M\rangle|n\rangle ),
    \end{aligned}
\label{eq:bas}
\end{equation}
with $\varepsilon = (-1)^{j_1+j_2-j_{12}+\ell}$. It imposes $\varepsilon=1$ for molecules in the same rovibrational level of the same electronic state \cite{li2019}.

The Hamiltonian \eqref{eq:hamilt} expressed in the basis \eqref{eq:bas} generates a set of close-coupled Schr\"odinger equations which are numerically solved \cite{sm} with a log-derivative method \cite{johnson1973,tuvi1993}, over the interval $\left[R_{min};R_{max} \right]=\left[10\,\textrm{a.u.};10000\,\textrm{a.u.} \right]$. We assume the scattering flux toward short distances is fully absorbed as in \cite{wang2015}. Such an approach has been successfully applied to simulate the observed reactive or non-reactive ultracold molecular collisions \cite{ospelkaus2010a,ni2010,ye2018}. The $S$ matrix is extracted at $R_{max}$, and rate coefficients are obtained.

We consider two $^{23}$Na$^{87}$Rb molecules prepared in their absolute ground state $|e_1 = X, j_1 = 0, p_1 = 1, e_2 = X, j_2 = 0, p_2 = 1, j_{12} = 0, \ell, J = \ell, M = 0\rangle |n=0\rangle$. In typical experimental conditions \cite{guo2016}, the temperature $T=E_{col}/k_B \approx 400$~nK only involves $\ell=0,2$. In the field-free case, the states correlated to the $(j_X=0,j_X=0)$ asymptote are restricted to $J=0$ and $J=2$. The DDI couples different $j_i$ and $\ell$ such that $|j'_i-j_i|=1$, $p_ip'_i=-1$, and $|\ell'-\ell|=0,2$. The long-range adiabatic PECs resulting from the diagonalization of the field-free Hamiltonian (Eq.\eqref{eq:hamilt} without $\hat{H}_f$ and $\hat{H}_I$) in the basis of Eq.\eqref{eq:bas} (without $|n=0\rangle$) are presented in the SM \cite{sm}. They are calculated for $j_i\in [0,j_\textrm{max}]$ and $\ell\in[0,\ell_\textrm{max}]$, where $j_\textrm{max}=4$ and $\ell_\textrm{max}=4$ to ensure the convergence of the rate coefficients.

\begin{figure}[!t]\centering
\resizebox{0.45\textwidth}{!}{%
  \includegraphics{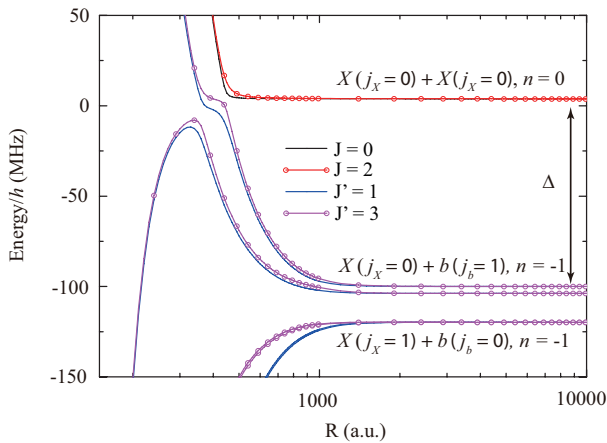}
}
    \caption{(Color online) The dressed adiabatic long-range PECs of $^{23}$Na$^{87}$Rb-$^{23}$Na$^{87}$Rb for $\Delta=100$~MHz and $\Omega=10$~MHz, in linear polarization. The Condon point (not displayed here) lies around $R_C = 400$~a.u.. The curves are labeled with their dominant $J$ character at large distances. Note that the asymptotic spacing between the $(j_X=0)+(j_X=0)$ and $(j_X=0)+(j_b=1)$ asymptotes is slightly larger (by 7.7~MHz) than $\Delta$ due to the presence of the stationary laser field \cite{napolitano1997}. For the same reason, the $(j_X=0)+(j_b=1)$ asymptote is split by 3.85~MHz as all the relevant states do not interact in the same way due to the laser.}
\label{fig:pecs}
\end{figure}

In Fig.~\ref{fig:pecs} we present the $^{23}$Na$^{87}$Rb-$^{23}$Na$^{87}$Rb dressed adiabatic long-range PECs in the presence of a linearly-polarized laser field with $\Delta = 100$~MHz and $\Omega = 10$~MHz. We note that the PECs are very similar in circularly-polarized light \cite{sm}. The states correlated to the $(j_X=0)+(j_X=0)$ asymptote with $n=0$ (with quantum numbers $J$ and $M$) are directly coupled to those correlated to the $(j_X=0)+(j_b=1)$ asymptote with $n=-1$ (with quantum numbers $J'$ and $M'$). Note that the  $(j_X=1)+(j_b=0)$ asymptote is distant by about $h\times$20~MHz from $(j_X=0)+(j_b=1)$ due to the almost identical values of the $b$ and $X$ rotational constants \cite{sm}, so that it must be included in the calculations. The DDI between these two sets of states induces the strong repulsive character of the upper manifold. Restricting the dressed-state approach to single photon transitions, the laser-induced couplings obey the electric-dipole selection rules, so that $|J'-J|=1$ and $M'=M$ (resp. $|J'-J|=0,1$ and $M'=M \pm 1$) for linearly or $\pi$ (resp. circularly or $\sigma^\pm$)-polarized light \cite{suominen1995}. We performed our study to the $M=0$ case, as the entrance channel is $(j_X=0)+(j_X=0)$ dominated by the $s$-wave ($\ell=0$) at ultracold energies.

The shielding possibility is clearly visible on Fig.~\ref{fig:pecs}. The dressed entrance channels correlated to $(j_X=0)+(j_X=0)$ now undergo avoided crossings with the repulsive channels correlated to $(j_X=0)+(j_b=1)$ so that the incoming flux is repelled, as demonstrated below. It is worth noting that the picture is similar for both circular and linear polarizations, in striking contrast with the molecular mw-shielding \cite{lassabliere2018,karman2018}, or with the atomic case \cite{marcassa1994}. Indeed the permutation symmetry selection rules allow the $d$-wave to be coupled to $(j_X=0)+(j_b=1)$ states not only in circular polarization but also in linear one.

\begin{figure}[!t]\centering
\resizebox{0.45\textwidth}{!}{%
  \includegraphics{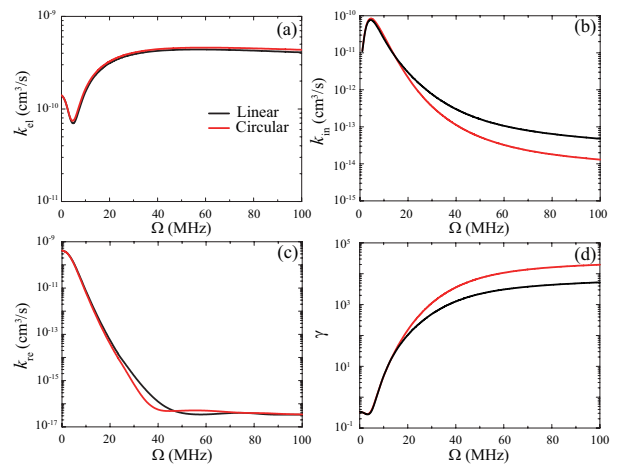}
}
    \caption{(Color online) Variation with the Rabi frequency $\Omega$ of the rate coefficients $k_\mathrm{el}$, $k_\mathrm{in}$, $k_\mathrm{re}$ for (a) elastic, (b) photoinduced inelastic, and (c) reactive collisions, respectively, and (d) of the shielding efficiency $\gamma$, for both linear and circular polarizations. The collision energy is $k_B \times$400~$n$K and the laser detuning $\Delta = 100$~MHz.}
    \label{fig:rate}
\end{figure}

The efficiency of the proposed scheme requires that the elastic collision rate $k_\mathrm{el}$ induced by the repulsive PECs in the entrance channel dominate over the loss rates, namely $k_\mathrm{re}$ for the short-range reactive collisions, and $k_\mathrm{in}$ for the photoinduced inelastic collisions $(j_X=0)+(j_X=0) \to (j_X=0)+(j_b=1)$. We look for conditions maximizing the ratio $\gamma = k_\mathrm{el} / (k_\mathrm{re} + k_\mathrm{in})$, often referred to as the ``good''-to-destructive collisional rate ratio \cite{gonzalez-martinez2017}.

In Fig.~\ref{fig:rate} we display the variation of the rate coefficients and $\gamma$ with the Rabi frequency $\Omega$ at fixed $\Delta = 100$~MHz for both polarizations. In the field-free case, the computed $k_\mathrm{re}=4.0 \times 10^{-10}$~cm$^3$.s$^{-1}$ is consistent with the experimental total loss rate ($4.5(2)\times 10^{-10}$~cm$^{3}$.s$^{-1}$) and the theoretical one ($3.8\times 10^{-10}$~cm$^{3}$.s$^{-1}$) reported in \cite{ye2018}. The computed value $\gamma\sim 0.3$ confirms the inefficiency of evaporative cooling, which requires $\gamma \gtrsim 1000$ \cite{ye2018}. The $k_\mathrm{re}$ rate drastically decreases with increasing $\Omega$, stabilizing to $k_\mathrm{re} \approx 3.0 \times 10^{-17}$~cm$^3$.s$^{-1}$ for $\Omega > 40$~MHz. After reaching a maximal value of about $10^{-10}$~cm$^{3}$.s$^{-1}$, the $k_\mathrm{in}$ value also strongly decreases, down to $5 \times 10^{-14}$~cm$^{3}$.s$^{-1}$ (resp. $3 \times 10^{-13}$~cm$^{3}$.s$^{-1}$) for circular (resp. linear) polarization at $\Omega = 50$~MHz. The $\gamma$ ratio is as high as 8000 and 2000 for circular and linear polarizations, respectively. We computed a transition dipole moment between $(j_X=0)$ and $(j_b=1)$ equal to 0.1918 a.u. \cite{sm} so that this regime is reached for a moderate intensity of about 12~W.cm$^{-2}$. It is worth recalling here that large $\gamma$ values imply dominant elastic collision rate, so that this shielding mechanism should allow an efficient evaporative cooling of the molecules.

\begin{figure}[!t]\centering
    \resizebox{0.4\textwidth}{!}{%
        \includegraphics{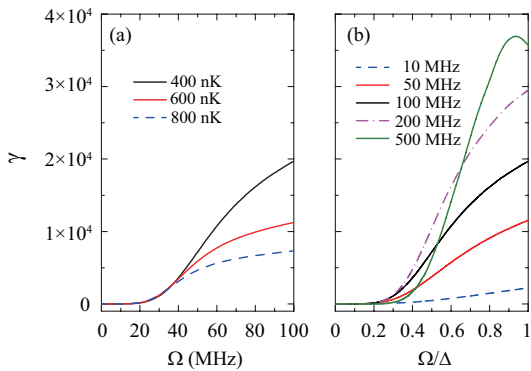}
    }
    \caption{(Color online) Variation of $\gamma$ in circularly-polarized light (a) with the Rabi frequency $\Omega$, at a fixed detuning $\Delta=100$~MHz, and  for three different temperatures $T = 400, 600, 800$~nK, and (b) with the ratio $\Omega/\Delta$ (kept $<1$), at a fixed temperature $T=400$~nK for five different detunings $\Delta =10, 50, 100, 200, 500$~MHz.}
\label{fig:gamma}
\end{figure}

The influence on $\gamma$ of the temperature $T=E_{col}/k_B$ and of the detuning $\Delta$ is illustrated in Fig.~\ref{fig:gamma} in circularly-polarized light as it appears more efficient (Fig.~\ref{fig:rate}). In panel (a), $\Delta=100$~MHz, and we see that $\gamma$ is insensitive to $T$ as long as the reactive rate dominates the inelastic one ($\Omega <20$~MHz), meaning that the avoided crossing is almost diabatic. When the reactive rate drops down abruptly, the inelastic rate reveals its sensitivity to the details of the avoided crossing: while $k_\mathrm{in}$ varies by order of magnitude when $\Omega'$ is multiplied by 5, $\gamma$ varies at most by a factor of 2 over the same range when changing the temperature by a factor of 2 (reflecting that $E\ll h\Omega$). In panel (b), we fixed $T=400$~nK, and $\gamma$ is displayed for convenience as a function of $\Omega/\Delta$ for various values of $\Delta$, thus all on the same scale. The restriction $\Omega/\Delta<1$ allows to keep valid and useful the representation of an avoided crossing (Fig. \ref{fig:pecs}). Again, no significant change is visible for $\Omega/\Delta<0.2$ (the reactive rate dominates) while the details of the avoided crossing are manifested above this value. At $\Delta = 500$~MHz, $\gamma$ reaches a maximum and then decreases, which is caused by the enhancement of photoinduced inelastic collision from the ground to the excited states. Large $\Delta$ values seem beneficial for shielding, but thus implies large $\Omega$ values, which may be problematic in terms of the power of light sources.

As noted in previous works on atoms, the spontaneous emission (SE) from the excited state during the collision is the main potential limitation of the proposed shielding effect. If an excited $^{23}$Na$^{87}$Rb molecule spontaneously emits a photon, it leads to a large increase in kinetic energy, which is likely to damage the optical shielding process. In this respect the mw-based shielding is advantageous as SE is negligible for ground-state rotational levels. The choice of the $b$ state in the present proposal actually fulfills this criterion: the lifetime of the $j_b=1$ level is $\tau_{\gamma}=6.97 \mu$s \cite{sm}. The classical picture of Fig.\ref{fig:scheme} is useful in this matter. Following \cite{suominen1996c}, the duration $\tau_{tp}$ of the complex classical motion from the Condon point $R_C$ to the turning point $R_{tp}$ is given by $\int_{R_C}^{R_{tp}}\frac{dR'}{v(R',p)}$ with $v(R',p)$ the classical local velocity which depends on the initial momentum $p_0$ and local PEC. The SE probability $P_{SE}$ during one collision can be estimated as $P_{SE}=P_b e^{-\tau_{\gamma}/\tau_{tp}}$ where $P_b$ is the population in the $b$ excited state. This formula has been successfully employed in atomic trap loss studies and compared with quantum results of \cite{julienne1994,boesten1994}. Here we found $\tau_{tp}=0.79$~ns for $\Delta=10$~MHz down to $\tau_{tp}=1.2$~ps for $\Delta=500$~MHz, indeed negligible compared to $\tau_{\gamma}$.

We predict that the proposed shielding mechanism is also valid for all heteronuclear alkali-metal diatomic but LiNa and KRb which do not have a large enough PDM, and thus low DDI \cite{vexiau2015}. Assuming that the DDI is proportional to the product of the PDM of the colliding molecules, one can scale it along the series of species \cite{sm} and repeat the same calculations. We found very similar variations of $\gamma$, shifted toward lower or larger values of $\Omega$. This is illustrated in Table \ref{TabI}, where we see that similar shielding efficiency ($\gamma=1000$) can be obtained for experimentally acceptable laser intensities (except for LiK). Moreover, being very similar in terms of electronic structure, these species are all expected to possess a long-lived electronic $b$ state. Therefore evaporative cooling appears as experimentally feasible for all the reported species.

\begin{table}[]
    \centering
    \setlength{\tabcolsep}{6mm}{
    \begin{tabular}{@{}ccc@{}}
 \hline\hline
  Species             &  $\Omega_{\gamma=1000}$ (MHz)& $I_{\gamma=1000}$ (W.cm$^{-2}$)\\
  \hline
  $^7$Li$^{39}$K      &  225.1                    & 3294     \\
  $^7$Li$^{87}$Rb     &  53.8                     & 35     \\
  $^7$Li$^{133}$Cs    &  36.1                     & 14   \\
  $^{23}$Na$^{39}$K   &  92.8                     & 265    \\
  $^{23}$Na$^{87}$Rb  &  29.3                     & 6.3     \\
  $^{23}$Na$^{133}$Cs &  24.5                     & 2.0      \\
  $^{39}$K$^{133}$Cs  &  48.8                     & 14      \\
  $^{87}$Rb$^{133}$Cs &  101.2                    & 53      \\
   \hline\hline
    \end{tabular}}
    \caption{Estimation of $\Omega$ values and corresponding laser intensities to reach  $\gamma=1000$, at 400~nK, with circularly polarized light and $\Delta=100$~MHz (see text), for bosonic alkali-metal diatomics. The LiNa and KRb species are not listed as they do not have large enough DDI to allow such a large value of $\gamma$.}
    \label{TabI}
\end{table}


 Just like in \cite{karman2018,lassabliere2018}, the present proposal is formulated in free space. As noted above, we demonstrate in the SM that a trapping laser, and a static magnetic field with a sufficient magnitude do not alter the OS. However most ongoing experiments with dipolar particles are intended to exemplify anisotropic interactions when they are exposed to a static electric field. Here an electric field as weak as about 200~V/cm splits the closely-spaced $j_X=0+j_b=1$ and $j_X=1+j_b=0$ manifolds, and induces the mixing of sublevels within the same $M$ subspace. Our preliminary investigation indicates that the barrier used for OS surely exists for any field intensity due to the interplay of DDI within these two manifolds. We do expect a good OS efficiency even in the presence of an electric field, providing that the OS laser frequency is properly tuned. For instance, at 1.5~kV/cm, typical of ongoing experiments, the OS laser frequency could be changed by about 500~MHz compared to the field-free case. Further calculations including an electric field will be implemented in our next work.

 We acknowledge support from the BLUESHIELD project (ANR-14-CE34-0006 from ``Agence Nationale de la Recherche''), LABEX PALM (www.labex-palm.fr), DIM SIRTEQ (www.sirteq.org), and a PHC BALATON (Campus France, 41919RK, and T\'{e}T-Fr 2018.2.1.13). We thank E. Luc, G. Qu\'{e}m\'{e}ner,  M. Raoult, and D. Wang for stimulating discussions. Calculations have been performed at the computing center M\'esoLUM (LUMAT FR2764), and HPC resources from DNUM CCUB (Centre de Calcul de l'Universit\'e de Bourgogne).


\begin{thebibliography}{53}%
	\makeatletter
	\providecommand \@ifxundefined [1]{%
		\@ifx{#1\undefined}
	}%
	\providecommand \@ifnum [1]{%
		\ifnum #1\expandafter \@firstoftwo
		\else \expandafter \@secondoftwo
		\fi
	}%
	\providecommand \@ifx [1]{%
		\ifx #1\expandafter \@firstoftwo
		\else \expandafter \@secondoftwo
		\fi
	}%
	\providecommand \natexlab [1]{#1}%
	\providecommand \enquote  [1]{``#1''}%
	\providecommand \bibnamefont  [1]{#1}%
	\providecommand \bibfnamefont [1]{#1}%
	\providecommand \citenamefont [1]{#1}%
	\providecommand \href@noop [0]{\@secondoftwo}%
	\providecommand \href [0]{\begingroup \@sanitize@url \@href}%
	\providecommand \@href[1]{\@@startlink{#1}\@@href}%
	\providecommand \@@href[1]{\endgroup#1\@@endlink}%
	\providecommand \@sanitize@url [0]{\catcode `\\12\catcode `\$12\catcode
		`\&12\catcode `\#12\catcode `\^12\catcode `\_12\catcode `\%12\relax}%
	\providecommand \@@startlink[1]{}%
	\providecommand \@@endlink[0]{}%
	\providecommand \url  [0]{\begingroup\@sanitize@url \@url }%
	\providecommand \@url [1]{\endgroup\@href {#1}{\urlprefix }}%
	\providecommand \urlprefix  [0]{URL }%
	\providecommand \Eprint [0]{\href }%
	\providecommand \doibase [0]{http://dx.doi.org/}%
	\providecommand \selectlanguage [0]{\@gobble}%
	\providecommand \bibinfo  [0]{\@secondoftwo}%
	\providecommand \bibfield  [0]{\@secondoftwo}%
	\providecommand \translation [1]{[#1]}%
	\providecommand \BibitemOpen [0]{}%
	\providecommand \bibitemStop [0]{}%
	\providecommand \bibitemNoStop [0]{.\EOS\space}%
	\providecommand \EOS [0]{\spacefactor3000\relax}%
	\providecommand \BibitemShut  [1]{\csname bibitem#1\endcsname}%
	\let\auto@bib@innerbib\@empty
	\bibitem [{\citenamefont {Baranov}(2007)}]{baranov2008}%
	\BibitemOpen
	\bibfield  {author} {\bibinfo {author} {\bibfnamefont {M.~A.}\ \bibnamefont
			{Baranov}},\ }\href@noop {} {\bibfield  {journal} {\bibinfo  {journal} {Phys.
				Rep.}\ }\textbf {\bibinfo {volume} {464}},\ \bibinfo {pages} {71} (\bibinfo
		{year} {2007})}\BibitemShut {NoStop}%
	\bibitem [{\citenamefont {Bloch}\ \emph {et~al.}(2008)\citenamefont {Bloch},
		\citenamefont {Dalibard},\ and\ \citenamefont {Zwerger}}]{bloch2008}%
	\BibitemOpen
	\bibfield  {author} {\bibinfo {author} {\bibfnamefont {I.}~\bibnamefont
			{Bloch}}, \bibinfo {author} {\bibfnamefont {J.}~\bibnamefont {Dalibard}}, \
		and\ \bibinfo {author} {\bibfnamefont {W.}~\bibnamefont {Zwerger}},\
	}\href@noop {} {\bibfield  {journal} {\bibinfo  {journal} {Rev. Mod. Phys.}\
		}\textbf {\bibinfo {volume} {80}},\ \bibinfo {pages} {885} (\bibinfo {year}
		{2008})}\BibitemShut {NoStop}%
	\bibitem [{\citenamefont {Lahaye}\ \emph {et~al.}(2009)\citenamefont {Lahaye},
		\citenamefont {Menotti}, \citenamefont {Santos}, \citenamefont {Lewenstein},\
		and\ \citenamefont {Pfau}}]{lahaye2009}%
	\BibitemOpen
	\bibfield  {author} {\bibinfo {author} {\bibfnamefont {T.}~\bibnamefont
			{Lahaye}}, \bibinfo {author} {\bibfnamefont {C.}~\bibnamefont {Menotti}},
		\bibinfo {author} {\bibfnamefont {L.}~\bibnamefont {Santos}}, \bibinfo
		{author} {\bibfnamefont {M.}~\bibnamefont {Lewenstein}}, \ and\ \bibinfo
		{author} {\bibfnamefont {T.}~\bibnamefont {Pfau}},\ }\href@noop {} {\bibfield
		{journal} {\bibinfo  {journal} {Rep. Prog. Phys.}\ }\textbf {\bibinfo
			{volume} {72}},\ \bibinfo {pages} {126401} (\bibinfo {year}
		{2009})}\BibitemShut {NoStop}%
	\bibitem [{\citenamefont {Bloch}\ \emph {et~al.}(2005)\citenamefont {Bloch},
		\citenamefont {Dalibard},\ and\ \citenamefont {Nascimb\`ene}}]{bloch2012}%
	\BibitemOpen
	\bibfield  {author} {\bibinfo {author} {\bibfnamefont {I.}~\bibnamefont
			{Bloch}}, \bibinfo {author} {\bibfnamefont {J.}~\bibnamefont {Dalibard}}, \
		and\ \bibinfo {author} {\bibfnamefont {S.}~\bibnamefont {Nascimb\`ene}},\
	}\href@noop {} {\bibfield  {journal} {\bibinfo  {journal} {Nature Phys.}\
		}\textbf {\bibinfo {volume} {38}},\ \bibinfo {pages} {S629} (\bibinfo {year}
		{2005})}\BibitemShut {NoStop}%
	\bibitem [{\citenamefont {Schwerdfeger}(2009)}]{leshouches2009}%
	\BibitemOpen
	\bibfield  {author} {\bibinfo {author} {\bibfnamefont {P.}~\bibnamefont
			{Schwerdfeger}}\ }(\bibinfo  {publisher} {Oxford University Press},\ \bibinfo
	{address} {Amsterdam},\ \bibinfo {year} {2009})\BibitemShut {NoStop}%
	\bibitem [{\citenamefont {Doyle}\ \emph {et~al.}(2004)\citenamefont {Doyle},
		\citenamefont {Friedrich}, \citenamefont {Krems},\ and\ \citenamefont
		{Masnou-Seeuws}}]{doyle2004}%
	\BibitemOpen
	\bibfield  {author} {\bibinfo {author} {\bibfnamefont {J.}~\bibnamefont
			{Doyle}}, \bibinfo {author} {\bibfnamefont {B.}~\bibnamefont {Friedrich}},
		\bibinfo {author} {\bibfnamefont {R.}~\bibnamefont {Krems}}, \ and\ \bibinfo
		{author} {\bibfnamefont {F.}~\bibnamefont {Masnou-Seeuws}},\ }\href@noop {}
	{\bibfield  {journal} {\bibinfo  {journal} {Eur. Phys. J. D}\ }\textbf
		{\bibinfo {volume} {31}},\ \bibinfo {pages} {149} (\bibinfo {year}
		{2004})}\BibitemShut {NoStop}%
	\bibitem [{\citenamefont {Carr}\ and\ \citenamefont {Ye}(2009)}]{carr2009}%
	\BibitemOpen
	\bibfield  {author} {\bibinfo {author} {\bibfnamefont {L.~D.}\ \bibnamefont
			{Carr}}\ and\ \bibinfo {author} {\bibfnamefont {J.}~\bibnamefont {Ye}},\
	}\href@noop {} {\bibfield  {journal} {\bibinfo  {journal} {New J. Phys.}\
		}\textbf {\bibinfo {volume} {11}},\ \bibinfo {pages} {055009} (\bibinfo
		{year} {2009})}\BibitemShut {NoStop}%
	\bibitem [{\citenamefont {Dulieu}\ and\ \citenamefont
		{Gabbanini}(2009)}]{dulieu2009}%
	\BibitemOpen
	\bibfield  {author} {\bibinfo {author} {\bibfnamefont {O.}~\bibnamefont
			{Dulieu}}\ and\ \bibinfo {author} {\bibfnamefont {C.}~\bibnamefont
			{Gabbanini}},\ }\href@noop {} {\bibfield  {journal} {\bibinfo  {journal}
			{Rep. Prog. Phys.}\ }\textbf {\bibinfo {volume} {72}},\ \bibinfo {pages}
		{086401} (\bibinfo {year} {2009})}\BibitemShut {NoStop}%
	\bibitem [{\citenamefont {Qu{\'e}m{\'e}ner}\ and\ \citenamefont
		{Julienne}(2012)}]{quemener2012}%
	\BibitemOpen
	\bibfield  {author} {\bibinfo {author} {\bibfnamefont {G.}~\bibnamefont
			{Qu{\'e}m{\'e}ner}}\ and\ \bibinfo {author} {\bibfnamefont {P.~S.}\
			\bibnamefont {Julienne}},\ }\href@noop {} {\bibfield  {journal} {\bibinfo
			{journal} {Chem. Rev.}\ }\textbf {\bibinfo {volume} {112}},\ \bibinfo {pages}
		{4949} (\bibinfo {year} {2012})}\BibitemShut {NoStop}%
	\bibitem [{\citenamefont {Moses}\ \emph {et~al.}(2017)\citenamefont {Moses},
		\citenamefont {Covey}, \citenamefont {Miecnikowski}, \citenamefont {Jin},\
		and\ \citenamefont {Ye}}]{moses2017}%
	\BibitemOpen
	\bibfield  {author} {\bibinfo {author} {\bibfnamefont {S.~A.}\ \bibnamefont
			{Moses}}, \bibinfo {author} {\bibfnamefont {J.~P.}\ \bibnamefont {Covey}},
		\bibinfo {author} {\bibfnamefont {M.~T.}\ \bibnamefont {Miecnikowski}},
		\bibinfo {author} {\bibfnamefont {D.~S.}\ \bibnamefont {Jin}}, \ and\
		\bibinfo {author} {\bibfnamefont {J.}~\bibnamefont {Ye}},\ }\href@noop {}
	{\bibfield  {journal} {\bibinfo  {journal} {Nature Phys.}\ }\textbf {\bibinfo
			{volume} {13}},\ \bibinfo {pages} {13} (\bibinfo {year} {2017})}\BibitemShut
	{NoStop}%
	\bibitem [{\citenamefont {Bohn}\ \emph {et~al.}(2017)\citenamefont {Bohn},
		\citenamefont {Rey},\ and\ \citenamefont {Ye}}]{bohn2017}%
	\BibitemOpen
	\bibfield  {author} {\bibinfo {author} {\bibfnamefont {J.~L.}\ \bibnamefont
			{Bohn}}, \bibinfo {author} {\bibfnamefont {A.~M.}\ \bibnamefont {Rey}}, \
		and\ \bibinfo {author} {\bibfnamefont {J.}~\bibnamefont {Ye}},\ }\href@noop
	{} {\bibfield  {journal} {\bibinfo  {journal} {Science}\ }\textbf {\bibinfo
			{volume} {357}},\ \bibinfo {pages} {1002} (\bibinfo {year}
		{2017})}\BibitemShut {NoStop}%
	\bibitem [{\citenamefont {Ospelkaus}\ \emph {et~al.}(2010)\citenamefont
		{Ospelkaus}, \citenamefont {Ni}, \citenamefont {Wang}, \citenamefont
		{de~Miranda}, \citenamefont {Neyenhuis}, \citenamefont {Qu\'em\'ener},
		\citenamefont {Julienne}, \citenamefont {Bohn}, \citenamefont {Jin},\ and\
		\citenamefont {Ye}}]{ospelkaus2010a}%
	\BibitemOpen
	\bibfield  {author} {\bibinfo {author} {\bibfnamefont {S.}~\bibnamefont
			{Ospelkaus}}, \bibinfo {author} {\bibfnamefont {K.-K.}\ \bibnamefont {Ni}},
		\bibinfo {author} {\bibfnamefont {D.}~\bibnamefont {Wang}}, \bibinfo {author}
		{\bibfnamefont {M.~H.~G.}\ \bibnamefont {de~Miranda}}, \bibinfo {author}
		{\bibfnamefont {B.}~\bibnamefont {Neyenhuis}}, \bibinfo {author}
		{\bibfnamefont {G.}~\bibnamefont {Qu\'em\'ener}}, \bibinfo {author}
		{\bibfnamefont {P.~S.}\ \bibnamefont {Julienne}}, \bibinfo {author}
		{\bibfnamefont {J.}~\bibnamefont {Bohn}}, \bibinfo {author} {\bibfnamefont
			{D.~S.}\ \bibnamefont {Jin}}, \ and\ \bibinfo {author} {\bibfnamefont
			{J.}~\bibnamefont {Ye}},\ }\href@noop {} {\bibfield  {journal} {\bibinfo
			{journal} {Science}\ }\textbf {\bibinfo {volume} {327}},\ \bibinfo {pages}
		{853} (\bibinfo {year} {2010})}\BibitemShut {NoStop}%
	\bibitem [{\citenamefont {Ni}\ \emph {et~al.}(2010)\citenamefont {Ni},
		\citenamefont {Ospelkaus}, \citenamefont {Wang}, \citenamefont
		{Qu\'em\'ener}, \citenamefont {Neyenhuis}, \citenamefont {de~Miranda},
		\citenamefont {Bohn}, \citenamefont {Ye},\ and\ \citenamefont
		{Jin}}]{ni2010}%
	\BibitemOpen
	\bibfield  {author} {\bibinfo {author} {\bibfnamefont {K.-K.}\ \bibnamefont
			{Ni}}, \bibinfo {author} {\bibfnamefont {S.}~\bibnamefont {Ospelkaus}},
		\bibinfo {author} {\bibfnamefont {D.}~\bibnamefont {Wang}}, \bibinfo {author}
		{\bibfnamefont {G.}~\bibnamefont {Qu\'em\'ener}}, \bibinfo {author}
		{\bibfnamefont {B.}~\bibnamefont {Neyenhuis}}, \bibinfo {author}
		{\bibfnamefont {M.~H.~G.}\ \bibnamefont {de~Miranda}}, \bibinfo {author}
		{\bibfnamefont {J.~L.}\ \bibnamefont {Bohn}}, \bibinfo {author}
		{\bibfnamefont {J.}~\bibnamefont {Ye}}, \ and\ \bibinfo {author}
		{\bibfnamefont {D.~S.}\ \bibnamefont {Jin}},\ }\href@noop {} {\bibfield
		{journal} {\bibinfo  {journal} {Nature}\ }\textbf {\bibinfo {volume} {464}},\
		\bibinfo {pages} {1324} (\bibinfo {year} {2010})}\BibitemShut {NoStop}%
	\bibitem [{\citenamefont {Takekoshi}\ \emph {et~al.}(2014)\citenamefont
		{Takekoshi}, \citenamefont {Reichs\"ollner}, \citenamefont {Schindewolf},
		\citenamefont {Hutson}, \citenamefont {LeSueur}, \citenamefont {Dulieu},
		\citenamefont {Ferlaino}, \citenamefont {Grimm},\ and\ \citenamefont
		{N\"agerl}}]{takekoshi2014}%
	\BibitemOpen
	\bibfield  {author} {\bibinfo {author} {\bibfnamefont {T.}~\bibnamefont
			{Takekoshi}}, \bibinfo {author} {\bibfnamefont {L.}~\bibnamefont
			{Reichs\"ollner}}, \bibinfo {author} {\bibfnamefont {A.}~\bibnamefont
			{Schindewolf}}, \bibinfo {author} {\bibfnamefont {J.~M.}\ \bibnamefont
			{Hutson}}, \bibinfo {author} {\bibfnamefont {C.~R.}\ \bibnamefont
			{LeSueur}}, \bibinfo {author} {\bibfnamefont {O.}~\bibnamefont {Dulieu}},
		\bibinfo {author} {\bibfnamefont {F.}~\bibnamefont {Ferlaino}}, \bibinfo
		{author} {\bibfnamefont {R.}~\bibnamefont {Grimm}}, \ and\ \bibinfo {author}
		{\bibfnamefont {H.-C.}\ \bibnamefont {N\"agerl}},\ }\href@noop {} {\bibfield
		{journal} {\bibinfo  {journal} {Phys. Rev. Lett.}\ }\textbf {\bibinfo
			{volume} {113}},\ \bibinfo {pages} {205301} (\bibinfo {year}
		{2014})}\BibitemShut {NoStop}%
	\bibitem [{\citenamefont {Molony}\ \emph {et~al.}(2014)\citenamefont {Molony},
		\citenamefont {Gregory}, \citenamefont {Ji}, \citenamefont {Lu},
		\citenamefont {K\"oppinger}, \citenamefont {Le~Sueur}, \citenamefont
		{Blackley}, \citenamefont {Hutson},\ and\ \citenamefont
		{Cornish}}]{molony2014}%
	\BibitemOpen
	\bibfield  {author} {\bibinfo {author} {\bibfnamefont {P.~K.}\ \bibnamefont
			{Molony}}, \bibinfo {author} {\bibfnamefont {P.~D.}\ \bibnamefont {Gregory}},
		\bibinfo {author} {\bibfnamefont {Z.}~\bibnamefont {Ji}}, \bibinfo {author}
		{\bibfnamefont {B.}~\bibnamefont {Lu}}, \bibinfo {author} {\bibfnamefont
			{M.~P.}\ \bibnamefont {K\"oppinger}}, \bibinfo {author} {\bibfnamefont
			{C.~R.}\ \bibnamefont {Le~Sueur}}, \bibinfo {author} {\bibfnamefont {C.~L.}\
			\bibnamefont {Blackley}}, \bibinfo {author} {\bibfnamefont {J.~M.}\
			\bibnamefont {Hutson}}, \ and\ \bibinfo {author} {\bibfnamefont {S.~L.}\
			\bibnamefont {Cornish}},\ }\href@noop {} {\bibfield  {journal} {\bibinfo
			{journal} {Phys. Rev. Lett.}\ }\textbf {\bibinfo {volume} {113}},\ \bibinfo
		{pages} {255301} (\bibinfo {year} {2014})}\BibitemShut {NoStop}%
	\bibitem [{\citenamefont {Gregory}\ \emph {et~al.}(2019)\citenamefont
		{Gregory}, \citenamefont {Frye}, \citenamefont {Blackmore}, \citenamefont
		{Bridge}, \citenamefont {Sawant}, \citenamefont {Hutson}, ,\ and\
		\citenamefont {Cornish}}]{gregory2019}%
	\BibitemOpen
	\bibfield  {author} {\bibinfo {author} {\bibfnamefont {P.~D.}\ \bibnamefont
			{Gregory}}, \bibinfo {author} {\bibfnamefont {M.~D.}\ \bibnamefont {Frye}},
		\bibinfo {author} {\bibfnamefont {J.~A.}\ \bibnamefont {Blackmore}}, \bibinfo
		{author} {\bibfnamefont {E.~M.}\ \bibnamefont {Bridge}}, \bibinfo {author}
		{\bibfnamefont {R.}~\bibnamefont {Sawant}}, \bibinfo {author} {\bibfnamefont
			{J.~M.}\ \bibnamefont {Hutson}}, , \ and\ \bibinfo {author} {\bibfnamefont
			{S.~L.}\ \bibnamefont {Cornish}},\ }\href@noop {} {\bibfield  {journal}
		{\bibinfo  {journal} {Nature Comm.}\ }\textbf {\bibinfo {volume} {10}},\
		\bibinfo {pages} {3104} (\bibinfo {year} {2019})}\BibitemShut {NoStop}%
	\bibitem [{\citenamefont {Guo}\ \emph {et~al.}(2016)\citenamefont {Guo},
		\citenamefont {Zhu}, \citenamefont {Lu}, \citenamefont {Ye}, \citenamefont
		{Wang}, \citenamefont {Vexiau}, \citenamefont {Bouloufa-Maafa}, \citenamefont
		{Qu\'em\'ener}, \citenamefont {Dulieu},\ and\ \citenamefont
		{Wang}}]{guo2016}%
	\BibitemOpen
	\bibfield  {author} {\bibinfo {author} {\bibfnamefont {M.}~\bibnamefont
			{Guo}}, \bibinfo {author} {\bibfnamefont {B.}~\bibnamefont {Zhu}}, \bibinfo
		{author} {\bibfnamefont {B.}~\bibnamefont {Lu}}, \bibinfo {author}
		{\bibfnamefont {X.}~\bibnamefont {Ye}}, \bibinfo {author} {\bibfnamefont
			{F.}~\bibnamefont {Wang}}, \bibinfo {author} {\bibfnamefont {R.}~\bibnamefont
			{Vexiau}}, \bibinfo {author} {\bibfnamefont {N.}~\bibnamefont
			{Bouloufa-Maafa}}, \bibinfo {author} {\bibfnamefont {G.}~\bibnamefont
			{Qu\'em\'ener}}, \bibinfo {author} {\bibfnamefont {O.}~\bibnamefont
			{Dulieu}}, \ and\ \bibinfo {author} {\bibfnamefont {D.}~\bibnamefont
			{Wang}},\ }\href@noop {} {\bibfield  {journal} {\bibinfo  {journal} {Phys.
				Rev. Lett.}\ }\textbf {\bibinfo {volume} {116}},\ \bibinfo {pages} {205303}
		(\bibinfo {year} {2016})}\BibitemShut {NoStop}%
	\bibitem [{\citenamefont {Park}\ \emph {et~al.}(2015)\citenamefont {Park},
		\citenamefont {Will},\ and\ \citenamefont {Zwierlein}}]{park2015}%
	\BibitemOpen
	\bibfield  {author} {\bibinfo {author} {\bibfnamefont {J.~W.}\ \bibnamefont
			{Park}}, \bibinfo {author} {\bibfnamefont {S.~A.}\ \bibnamefont {Will}}, \
		and\ \bibinfo {author} {\bibfnamefont {M.~W.}\ \bibnamefont {Zwierlein}},\
	}\href@noop {} {\bibfield  {journal} {\bibinfo  {journal} {Phys. Rev. Lett.}\
		}\textbf {\bibinfo {volume} {114}},\ \bibinfo {pages} {205302} (\bibinfo
		{year} {2015})}\BibitemShut {NoStop}%
	\bibitem [{\citenamefont {See\ss{}elberg}\ \emph {et~al.}(2018)\citenamefont
		{See\ss{}elberg}, \citenamefont {Buchheim}, \citenamefont {Lu}, \citenamefont
		{Schneider}, \citenamefont {Luo}, \citenamefont {Tiemann}, \citenamefont
		{Bloch},\ and\ \citenamefont {Gohle}}]{seesselberg2018}%
	\BibitemOpen
	\bibfield  {author} {\bibinfo {author} {\bibfnamefont {F.}~\bibnamefont
			{See\ss{}elberg}}, \bibinfo {author} {\bibfnamefont {N.}~\bibnamefont
			{Buchheim}}, \bibinfo {author} {\bibfnamefont {Z.-K.}\ \bibnamefont {Lu}},
		\bibinfo {author} {\bibfnamefont {T.}~\bibnamefont {Schneider}}, \bibinfo
		{author} {\bibfnamefont {X.-Y.}\ \bibnamefont {Luo}}, \bibinfo {author}
		{\bibfnamefont {E.}~\bibnamefont {Tiemann}}, \bibinfo {author} {\bibfnamefont
			{I.}~\bibnamefont {Bloch}}, \ and\ \bibinfo {author} {\bibfnamefont
			{C.}~\bibnamefont {Gohle}},\ }\href@noop {} {\bibfield  {journal} {\bibinfo
			{journal} {Phys. Rev. A}\ }\textbf {\bibinfo {volume} {97}},\ \bibinfo
		{pages} {013405} (\bibinfo {year} {2018})}\BibitemShut {NoStop}%
	\bibitem [{\citenamefont {\.Zuchowski}\ \emph {et~al.}(2010)\citenamefont
		{\.Zuchowski}, \ and\ \citenamefont
		{Hutson}}]{zuchowski2010}%
	\BibitemOpen
	\bibfield  {author} {\bibinfo {author} {\bibfnamefont {P.~S.}\ \bibnamefont
			{\.Zuchowski}}, \ and\ \bibinfo {author} {\bibfnamefont {J.~M.}\ \bibnamefont
			{Hutson}},\ }\href@noop {} {\bibfield  {journal} {\bibinfo  {journal} {Phys.
				Rev. A}\ }\textbf {\bibinfo {volume} {81}},\ \bibinfo {pages} {060703(R)}
		(\bibinfo {year} {2010})}\BibitemShut {NoStop}%
	\bibitem [{\citenamefont {Ye}\ \emph {et~al.}(2018{\natexlab{a}})\citenamefont
		{Ye}, \citenamefont {Guo}, \citenamefont {Gonz{\'a}lez-Mart{\'\i}nez},
		\citenamefont {Qu{\'e}m{\'e}ner},\ and\ \citenamefont {Wang}}]{guo2018}%
	\BibitemOpen
	\bibfield  {author} {\bibinfo {author} {\bibfnamefont {X.}~\bibnamefont
			{Ye}}, \bibinfo {author} {\bibfnamefont {M.}~\bibnamefont {Guo}}, \bibinfo
		{author} {\bibfnamefont {M.~L.}\ \bibnamefont {Gonz{\'a}lez-Mart{\'\i}nez}},
		\bibinfo {author} {\bibfnamefont {G.}~\bibnamefont {Qu{\'e}m{\'e}ner}}, \
		and\ \bibinfo {author} {\bibfnamefont {D.}~\bibnamefont {Wang}},\ }\href@noop
	{} {\bibfield  {journal} {\bibinfo  {journal} {Science Advances}\ }\textbf
		{\bibinfo {volume} {4}} (\bibinfo {year} {2018}{\natexlab{a}})}\BibitemShut
	{NoStop}%
	\bibitem [{\citenamefont {Hu}\ \emph {et~al.}(2019)\citenamefont {Hu},
		\citenamefont {Liu}, \citenamefont {Grimes}, \citenamefont {Lin},
		\citenamefont {Gheorghe}, \citenamefont {Vexiau}, \citenamefont
		{Bouloufa-Maafa}, \citenamefont {Dulieu}, \citenamefont {Rosenband},\ and\
		\citenamefont {Ni}}]{hu2019}%
	\BibitemOpen
	\bibfield  {author} {\bibinfo {author} {\bibfnamefont {M.-G.}\ \bibnamefont
			{Hu}}, \bibinfo {author} {\bibfnamefont {Y.}~\bibnamefont {Liu}}, \bibinfo
		{author} {\bibfnamefont {D.~D.}\ \bibnamefont {Grimes}}, \bibinfo {author}
		{\bibfnamefont {Y.-W.}\ \bibnamefont {Lin}}, \bibinfo {author} {\bibfnamefont
			{A.~H.}\ \bibnamefont {Gheorghe}}, \bibinfo {author} {\bibfnamefont
			{R.}~\bibnamefont {Vexiau}}, \bibinfo {author} {\bibfnamefont
			{N.}~\bibnamefont {Bouloufa-Maafa}}, \bibinfo {author} {\bibfnamefont
			{O.}~\bibnamefont {Dulieu}}, \bibinfo {author} {\bibfnamefont
			{T.}~\bibnamefont {Rosenband}}, \ and\ \bibinfo {author} {\bibfnamefont
			{K.-K.}\ \bibnamefont {Ni}},\ }\href@noop {} {\bibfield  {journal} {\bibinfo
			{journal} {Science}\ }\textbf {\bibinfo {volume} {366}},\ \bibinfo {pages}
		{1111} (\bibinfo {year} {2019})}\BibitemShut {NoStop}%
	\bibitem [{\citenamefont {Liu}\ \emph {et~al.}(2020)\citenamefont {Liu},
		\citenamefont {Hu}, \citenamefont {Nichols}, \citenamefont {Grimes},
		\citenamefont {Karman}, \citenamefont {Guo},\ and\ \citenamefont
		{Ni}}]{liu2020}%
	\BibitemOpen
	\bibfield  {author} {\bibinfo {author} {\bibfnamefont {Y.}~\bibnamefont
			{Liu}}, \bibinfo {author} {\bibfnamefont {M.-G.}\ \bibnamefont {Hu}},
		\bibinfo {author} {\bibfnamefont {M.~A.}\ \bibnamefont {Nichols}}, \bibinfo
		{author} {\bibfnamefont {D.~D.}\ \bibnamefont {Grimes}}, \bibinfo {author}
		{\bibfnamefont {T.}~\bibnamefont {Karman}}, \bibinfo {author} {\bibfnamefont
			{H.}~\bibnamefont {Guo}}, \ and\ \bibinfo {author} {\bibfnamefont {K.-K.}\
			\bibnamefont {Ni}},\ }\href@noop {} {\bibfield  {journal} {\bibinfo
			{journal} {Nature Phys.}\ ,\ \bibinfo {pages}
			{https://doi.org/10.1038/s41567}} (\bibinfo {year} {2020})}\BibitemShut
	{NoStop}%
	\bibitem [{\citenamefont {Christianen}\ \emph {et~al.}(2019)\citenamefont
		{Christianen}, \citenamefont {Zwierlein}, \citenamefont {Groenenboom},\ and\
		\citenamefont {Karman}}]{christianen2019}%
	\BibitemOpen
	\bibfield  {author} {\bibinfo {author} {\bibfnamefont {A.}~\bibnamefont
			{Christianen}}, \bibinfo {author} {\bibfnamefont {M.~W.}\ \bibnamefont
			{Zwierlein}}, \bibinfo {author} {\bibfnamefont {G.~C.}\ \bibnamefont
			{Groenenboom}}, \ and\ \bibinfo {author} {\bibfnamefont {T.}~\bibnamefont
			{Karman}},\ }\href@noop {} {\bibfield  {journal} {\bibinfo  {journal} {Phys.
				Rev. Lett.}\ }\textbf {\bibinfo {volume} {123}},\ \bibinfo {pages} {123402}
		(\bibinfo {year} {2019})}\BibitemShut {NoStop}%
	\bibitem [{\citenamefont {Gregory}\ \emph {et~al.}(2020)\citenamefont
		{Gregory}, \citenamefont {Blackmore}, \citenamefont {Bromley},\ and\
		\citenamefont {Cornish}}]{gregory2020}%
	\BibitemOpen
	\bibfield  {author} {\bibinfo {author} {\bibfnamefont {P.~D.}\ \bibnamefont
			{Gregory}}, \bibinfo {author} {\bibfnamefont {J.~A.}\ \bibnamefont
			{Blackmore}}, \bibinfo {author} {\bibfnamefont {S.~L.}\ \bibnamefont
			{Bromley}}, \ and\ \bibinfo {author} {\bibfnamefont {S.~L.}\ \bibnamefont
			{Cornish}},\ }\href@noop {} {\bibfield  {journal} {\bibinfo  {journal} {Phys.
				Rev. Lett.}\ }\textbf {\bibinfo {volume} {124}},\ \bibinfo {pages} {163402}
		(\bibinfo {year} {2020})}\BibitemShut {NoStop}%
	\bibitem [{\citenamefont {Karman}\ and\ \citenamefont
		{Hutson}(2018)}]{karman2018}%
	\BibitemOpen
	\bibfield  {author} {\bibinfo {author} {\bibfnamefont {T.}~\bibnamefont
			{Karman}}\ and\ \bibinfo {author} {\bibfnamefont {J.~M.}\ \bibnamefont
			{Hutson}},\ }\href@noop {} {\bibfield  {journal} {\bibinfo  {journal} {Phys.
				Rev. Lett.}\ }\textbf {\bibinfo {volume} {121}},\ \bibinfo {pages} {163401}
		(\bibinfo {year} {2018})}\BibitemShut {NoStop}%
	\bibitem [{\citenamefont {Lassabli\`ere}\ and\ \citenamefont
		{Qu\'em\'ener}(2018)}]{lassabliere2018}%
	\BibitemOpen
	\bibfield  {author} {\bibinfo {author} {\bibfnamefont {L.}~\bibnamefont
			{Lassabli\`ere}}\ and\ \bibinfo {author} {\bibfnamefont {G.}~\bibnamefont
			{Qu\'em\'ener}},\ }\href@noop {} {\bibfield  {journal} {\bibinfo  {journal}
			{Phys. Rev. Lett.}\ }\textbf {\bibinfo {volume} {121}},\ \bibinfo {pages}
		{163402} (\bibinfo {year} {2018})}\BibitemShut {NoStop}%
	\bibitem [{\citenamefont {Avdeenkov}(2015)}]{avdeenkov2015}%
	\BibitemOpen
	\bibfield  {author} {\bibinfo {author} {\bibfnamefont {A.~V.}\ \bibnamefont
			{Avdeenkov}},\ }\href@noop {} {\bibfield  {journal} {\bibinfo  {journal} {New
				J. Phys.}\ }\textbf {\bibinfo {volume} {17}},\ \bibinfo {pages} {045025}
		(\bibinfo {year} {2015})}\BibitemShut {NoStop}%
	\bibitem [{\citenamefont {Karman}(2020)}]{karman2020}%
	\BibitemOpen
	\bibfield  {author} {\bibinfo {author} {\bibfnamefont {T.}~\bibnamefont
			{Karman}},\ }\href@noop {} {\bibfield  {journal} {\bibinfo  {journal} {Phys.
				Rev. A}\ }\textbf {\bibinfo {volume} {101}},\ \bibinfo {pages} {042702}
		(\bibinfo {year} {2020})}\BibitemShut {NoStop}%
	\bibitem [{\citenamefont {Napolitano}\ \emph {et~al.}(1997)\citenamefont
		{Napolitano}, \citenamefont {Weiner},\ and\ \citenamefont
		{Julienne}}]{napolitano1997}%
	\BibitemOpen
	\bibfield  {author} {\bibinfo {author} {\bibfnamefont {R.}~\bibnamefont
			{Napolitano}}, \bibinfo {author} {\bibfnamefont {J.}~\bibnamefont {Weiner}},
		\ and\ \bibinfo {author} {\bibfnamefont {P.~S.}\ \bibnamefont {Julienne}},\
	}\href@noop {} {\bibfield  {journal} {\bibinfo  {journal} {Phys. Rev. A}\
		}\textbf {\bibinfo {volume} {55}},\ \bibinfo {pages} {1191} (\bibinfo {year}
		{1997})}\BibitemShut {NoStop}%
	\bibitem [{\citenamefont {Bali}\ \emph {et~al.}(1994)\citenamefont {Bali},
		\citenamefont {Hoffmann},\ and\ \citenamefont {Walker}}]{bali1994}%
	\BibitemOpen
	\bibfield  {author} {\bibinfo {author} {\bibfnamefont {S.}~\bibnamefont
			{Bali}}, \bibinfo {author} {\bibfnamefont {D.}~\bibnamefont {Hoffmann}}, \
		and\ \bibinfo {author} {\bibfnamefont {T.}~\bibnamefont {Walker}},\
	}\href@noop {} {\bibfield  {journal} {\bibinfo  {journal} {Europhys. Lett.}\
		}\textbf {\bibinfo {volume} {27}},\ \bibinfo {pages} {273} (\bibinfo {year}
		{1994})}\BibitemShut {NoStop}%
	\bibitem [{\citenamefont {Marcassa}\ \emph {et~al.}(1994)\citenamefont
		{Marcassa}, \citenamefont {Muniz}, \citenamefont {de~Queiroz}, \citenamefont
		{Zilio}, \citenamefont {Bagnato}, \citenamefont {Weiner}, \citenamefont
		{Julienne},\ and\ \citenamefont {Suominen}}]{marcassa1994}%
	\BibitemOpen
	\bibfield  {author} {\bibinfo {author} {\bibfnamefont {L.}~\bibnamefont
			{Marcassa}}, \bibinfo {author} {\bibfnamefont {S.}~\bibnamefont {Muniz}},
		\bibinfo {author} {\bibfnamefont {E.}~\bibnamefont {de~Queiroz}}, \bibinfo
		{author} {\bibfnamefont {S.}~\bibnamefont {Zilio}}, \bibinfo {author}
		{\bibfnamefont {V.}~\bibnamefont {Bagnato}}, \bibinfo {author} {\bibfnamefont
			{J.}~\bibnamefont {Weiner}}, \bibinfo {author} {\bibfnamefont {P.~S.}\
			\bibnamefont {Julienne}}, \ and\ \bibinfo {author} {\bibfnamefont {K.~A.}\
			\bibnamefont {Suominen}},\ }\href@noop {} {\bibfield  {journal} {\bibinfo
			{journal} {Phys. Rev. Lett.}\ }\textbf {\bibinfo {volume} {73}},\ \bibinfo
		{pages} {1911} (\bibinfo {year} {1994})}\BibitemShut {NoStop}%
	\bibitem [{\citenamefont {Suominen}\ \emph
		{et~al.}(1996{\natexlab{a}})\citenamefont {Suominen}, \citenamefont
		{Burnett}, \citenamefont {Julienne}, \citenamefont {Walhout}, \citenamefont
		{Sterr}, \citenamefont {Orzel}, \citenamefont {Hoogerland},\ and\
		\citenamefont {Rolston}}]{suominen1996a}%
	\BibitemOpen
	\bibfield  {author} {\bibinfo {author} {\bibfnamefont {K.-A.}\ \bibnamefont
			{Suominen}}, \bibinfo {author} {\bibfnamefont {K.}~\bibnamefont {Burnett}},
		\bibinfo {author} {\bibfnamefont {P.~S.}\ \bibnamefont {Julienne}}, \bibinfo
		{author} {\bibfnamefont {M.}~\bibnamefont {Walhout}}, \bibinfo {author}
		{\bibfnamefont {U.}~\bibnamefont {Sterr}}, \bibinfo {author} {\bibfnamefont
			{C.}~\bibnamefont {Orzel}}, \bibinfo {author} {\bibfnamefont
			{M.}~\bibnamefont {Hoogerland}}, \ and\ \bibinfo {author} {\bibfnamefont
			{S.~L.}\ \bibnamefont {Rolston}},\ }\href@noop {} {\bibfield  {journal}
		{\bibinfo  {journal} {Phys. Rev. A}\ }\textbf {\bibinfo {volume} {53}},\
		\bibinfo {pages} {1678} (\bibinfo {year} {1996}{\natexlab{a}})}\BibitemShut
	{NoStop}%
	\bibitem [{\citenamefont {Hoffmann}\ \emph {et~al.}(1996)\citenamefont
		{Hoffmann}, \citenamefont {Bali},\ and\ \citenamefont
		{Walker}}]{hoffmann1996}%
	\BibitemOpen
	\bibfield  {author} {\bibinfo {author} {\bibfnamefont {D.}~\bibnamefont
			{Hoffmann}}, \bibinfo {author} {\bibfnamefont {S.}~\bibnamefont {Bali}}, \
		and\ \bibinfo {author} {\bibfnamefont {T.}~\bibnamefont {Walker}},\
	}\href@noop {} {\bibfield  {journal} {\bibinfo  {journal} {Phys. Rev. A}\
		}\textbf {\bibinfo {volume} {54}},\ \bibinfo {pages} {R1030} (\bibinfo {year}
		{1996})}\BibitemShut {NoStop}%
	\bibitem [{\citenamefont {Zilio}\ \emph {et~al.}(1996)\citenamefont {Zilio},
		\citenamefont {Marcassa}, \citenamefont {Muniz}, \citenamefont {Horowicz},
		\citenamefont {Bagnato}, \citenamefont {Napolitano}, \citenamefont {Weiner},\
		and\ \citenamefont {Julienne}}]{zilio1996}%
	\BibitemOpen
	\bibfield  {author} {\bibinfo {author} {\bibfnamefont {S.~C.}~\bibnamefont
			{Zilio}}, \bibinfo {author} {\bibfnamefont {L.}~\bibnamefont {Marcassa}},
		\bibinfo {author} {\bibfnamefont {S.}~\bibnamefont {Muniz}}, \bibinfo
		{author} {\bibfnamefont {R.}~\bibnamefont {Horowicz}}, \bibinfo {author}
		{\bibfnamefont {V.}~\bibnamefont {Bagnato}}, \bibinfo {author} {\bibfnamefont
			{R.}~\bibnamefont {Napolitano}}, \bibinfo {author} {\bibfnamefont
			{J.}~\bibnamefont {Weiner}}, \ and\ \bibinfo {author} {\bibfnamefont {P.~S.}\
			\bibnamefont {Julienne}},\ }\href@noop {} {\bibfield  {journal} {\bibinfo
			{journal} {Phys. Rev. Lett.}\ }\textbf {\bibinfo {volume} {76}},\ \bibinfo
		{pages} {2033} (\bibinfo {year} {1996})}\BibitemShut {NoStop}%
	\bibitem [{\citenamefont {Muniz}\ \emph {et~al.}(1997)\citenamefont {Muniz},
		\citenamefont {Marcassa}, \citenamefont {Napolitano}, \citenamefont {Telles},
		\citenamefont {Weiner}, \citenamefont {Zilio},\ and\ \citenamefont
		{Bagnato}}]{muniz1997}%
	\BibitemOpen
	\bibfield  {author} {\bibinfo {author} {\bibfnamefont {S.~R.}\ \bibnamefont
			{Muniz}}, \bibinfo {author} {\bibfnamefont {L.~G.}\ \bibnamefont {Marcassa}},
		\bibinfo {author} {\bibfnamefont {R.}~\bibnamefont {Napolitano}}, \bibinfo
		{author} {\bibfnamefont {G.~D.}\ \bibnamefont {Telles}}, \bibinfo {author}
		{\bibfnamefont {J.}~\bibnamefont {Weiner}}, \bibinfo {author} {\bibfnamefont
			{S.~C.}\ \bibnamefont {Zilio}}, \ and\ \bibinfo {author} {\bibfnamefont
			{V.~S.}\ \bibnamefont {Bagnato}},\ }\href@noop {} {\bibfield  {journal}
		{\bibinfo  {journal} {Phys. Rev. A}\ }\textbf {\bibinfo {volume} {55}},\
		\bibinfo {pages} {4407} (\bibinfo {year} {1997})}\BibitemShut {NoStop}%
	\bibitem [{\citenamefont {Weiner}\ \emph {et~al.}(1999)\citenamefont {Weiner},
		\citenamefont {Bagnato}, \citenamefont {Zilio},\ and\ \citenamefont
		{Julienne}}]{weiner1999}%
	\BibitemOpen
	\bibfield  {author} {\bibinfo {author} {\bibfnamefont {J.}~\bibnamefont
			{Weiner}}, \bibinfo {author} {\bibfnamefont {V.~S.}\ \bibnamefont {Bagnato}},
		\bibinfo {author} {\bibfnamefont {S.~C.}\ \bibnamefont {Zilio}}, \ and\
		\bibinfo {author} {\bibfnamefont {P.~S.}\ \bibnamefont {Julienne}},\
	}\href@noop {} {\bibfield  {journal} {\bibinfo  {journal} {Rev. Mod. Phys.}\
		}\textbf {\bibinfo {volume} {71}},\ \bibinfo {pages} {1} (\bibinfo {year}
		{1999})}\BibitemShut {NoStop}%
	\bibitem [{\citenamefont {Suominen}\ \emph {et~al.}(1995)\citenamefont
		{Suominen}, \citenamefont {Holland}, \citenamefont {Burnett},\ and\
		\citenamefont {Julienne}}]{suominen1995}%
	\BibitemOpen
	\bibfield  {author} {\bibinfo {author} {\bibfnamefont {K.-A.}\ \bibnamefont
			{Suominen}}, \bibinfo {author} {\bibfnamefont {M.~J.}\ \bibnamefont
			{Holland}}, \bibinfo {author} {\bibfnamefont {K.}~\bibnamefont {Burnett}}, \
		and\ \bibinfo {author} {\bibfnamefont {P.}~\bibnamefont {Julienne}},\
	}\href@noop {} {\bibfield  {journal} {\bibinfo  {journal} {Phys. Rev. A}\
		}\textbf {\bibinfo {volume} {51}},\ \bibinfo {pages} {1446} (\bibinfo {year}
		{1995})}\BibitemShut {NoStop}%
	\bibitem [{\citenamefont {Suominen}\ \emph
		{et~al.}(1996{\natexlab{b}})\citenamefont {Suominen}, \citenamefont
		{Burnett},\ and\ \citenamefont {Julienne}}]{suominen1996b}%
	\BibitemOpen
	\bibfield  {author} {\bibinfo {author} {\bibfnamefont {K.~A.}\ \bibnamefont
			{Suominen}}, \bibinfo {author} {\bibfnamefont {K.}~\bibnamefont {Burnett}}, \
		and\ \bibinfo {author} {\bibfnamefont {P.~S.}\ \bibnamefont {Julienne}},\
	}\href@noop {} {\bibfield  {journal} {\bibinfo  {journal} {Phys. Rev. A}\
		}\textbf {\bibinfo {volume} {53}},\ \bibinfo {pages} {R1220} (\bibinfo {year}
		{1996}{\natexlab{b}})}\BibitemShut {NoStop}%
	\bibitem [{\citenamefont {Docenko}\ \emph {et~al.}(2007)\citenamefont
		{Docenko}, \citenamefont {Tamanis}, \citenamefont {Ferber}, \citenamefont
		{Pazyuk}, \citenamefont {Zaitsevskii}, \citenamefont {Stolyarov},
		\citenamefont {Pashov}, \citenamefont {Kn\"{o}ckel},\ and\ \citenamefont
		{Tiemann}}]{docenko2007}%
	\BibitemOpen
	\bibfield  {author} {\bibinfo {author} {\bibfnamefont {O.}~\bibnamefont
			{Docenko}}, \bibinfo {author} {\bibfnamefont {M.}~\bibnamefont {Tamanis}},
		\bibinfo {author} {\bibfnamefont {R.}~\bibnamefont {Ferber}}, \bibinfo
		{author} {\bibfnamefont {E.~A.}\ \bibnamefont {Pazyuk}}, \bibinfo {author}
		{\bibfnamefont {A.}~\bibnamefont {Zaitsevskii}}, \bibinfo {author}
		{\bibfnamefont {A.~V.}\ \bibnamefont {Stolyarov}}, \bibinfo {author}
		{\bibfnamefont {A.}~\bibnamefont {Pashov}}, \bibinfo {author} {\bibfnamefont
			{H.}~\bibnamefont {Kn\"{o}ckel}}, \ and\ \bibinfo {author} {\bibfnamefont
			{E.}~\bibnamefont {Tiemann}},\ }\href@noop {} {\bibfield  {journal} {\bibinfo
			{journal} {Phys. Rev. A}\ }\textbf {\bibinfo {volume} {75}},\ \bibinfo
		{pages} {042503} (\bibinfo {year} {2007})}\BibitemShut {NoStop}%
	\bibitem [{\citenamefont {Lepers}\ \emph {et~al.}(2013)\citenamefont {Lepers},
		\citenamefont {Vexiau}, \citenamefont {Aymar}, \citenamefont
		{Bouloufa-Maafa},\ and\ \citenamefont {Dulieu}}]{lepers2013}%
	\BibitemOpen
	\bibfield  {author} {\bibinfo {author} {\bibfnamefont {M.}~\bibnamefont
			{Lepers}}, \bibinfo {author} {\bibfnamefont {R.}~\bibnamefont {Vexiau}},
		\bibinfo {author} {\bibfnamefont {M.}~\bibnamefont {Aymar}}, \bibinfo
		{author} {\bibfnamefont {N.}~\bibnamefont {Bouloufa-Maafa}}, \ and\ \bibinfo
		{author} {\bibfnamefont {O.}~\bibnamefont {Dulieu}},\ }\href@noop {}
	{\bibfield  {journal} {\bibinfo  {journal} {Phys. Rev. A}\ }\textbf {\bibinfo
			{volume} {88}},\ \bibinfo {pages} {032709} (\bibinfo {year}
		{2013})}\BibitemShut {NoStop}%
	\bibitem [{\citenamefont {Vexiau}\ \emph {et~al.}(2015)\citenamefont {Vexiau},
		\citenamefont {Lepers}, \citenamefont {Aymar}, \citenamefont
		{Bouloufa-Maafa},\ and\ \citenamefont {Dulieu}}]{vexiau2015}%
	\BibitemOpen
	\bibfield  {author} {\bibinfo {author} {\bibfnamefont {R.}~\bibnamefont
			{Vexiau}}, \bibinfo {author} {\bibfnamefont {M.}~\bibnamefont {Lepers}},
		\bibinfo {author} {\bibfnamefont {M.}~\bibnamefont {Aymar}}, \bibinfo
		{author} {\bibfnamefont {N.}~\bibnamefont {Bouloufa-Maafa}}, \ and\ \bibinfo
		{author} {\bibfnamefont {O.}~\bibnamefont {Dulieu}},\ }\href@noop {}
	{\bibfield  {journal} {\bibinfo  {journal} {J. Chem. Phys.}\ }\textbf
		{\bibinfo {volume} {142}},\ \bibinfo {pages} {214303} (\bibinfo {year}
		{2015})}\BibitemShut {NoStop}%
	\bibitem [{\citenamefont {Lepers}\ and\ \citenamefont
		{Dulieu}(2018)}]{lepers2018}%
	\BibitemOpen
	\bibfield  {author} {\bibinfo {author} {\bibfnamefont {M.}~\bibnamefont
			{Lepers}}\ and\ \bibinfo {author} {\bibfnamefont {O.}~\bibnamefont
			{Dulieu}},\ }in\ \href@noop {} {\emph {\bibinfo {booktitle} {Cold Chemistry:
				Molecular Scattering and Reactivity Near Absolute Zero}}}\ (\bibinfo
	{publisher} {The Royal Society of Chemistry},\ \bibinfo {year} {2018})\ pp.\
	\bibinfo {pages} {150--202}\BibitemShut {NoStop}%
	\bibitem [{\citenamefont {Li}\ \emph {et~al.}(2019)\citenamefont {Li},
		\citenamefont {Qu\'em\'ener}, \citenamefont {Wyart}, \citenamefont {Dulieu},\
		and\ \citenamefont {Lepers}}]{li2019}%
	\BibitemOpen
	\bibfield  {author} {\bibinfo {author} {\bibfnamefont {H.}~\bibnamefont
			{Li}}, \bibinfo {author} {\bibfnamefont {G.}~\bibnamefont {Qu\'em\'ener}},
		\bibinfo {author} {\bibfnamefont {J.~F.}~\bibnamefont {Wyart}},
		\bibinfo {author} {\bibfnamefont {O.}~\bibnamefont {Dulieu}}, \ and\ \bibinfo
		{author} {\bibfnamefont {M.}~\bibnamefont {Lepers}},\ }\href@noop {}
	{\bibfield  {journal} {\bibinfo  {journal} {Phys. Rev. A}\ }\textbf {\bibinfo
			{volume} {100}},\ \bibinfo {pages} {042711} (\bibinfo {year}
		{2019})}\BibitemShut {NoStop}%
	\bibitem [{\citenamefont {Cohen-Tannoudji}\ \emph {et~al.}(1998)\citenamefont
		{Cohen-Tannoudji}, \citenamefont {Dupont-Roc},\ and\ \citenamefont
		{Grynberg}}]{cohen-tannoudji1998}%
	\BibitemOpen
	\bibfield  {author} {\bibinfo {author} {\bibfnamefont {C.}~\bibnamefont
			{Cohen-Tannoudji}}, \bibinfo {author} {\bibfnamefont {J.}~\bibnamefont
			{Dupont-Roc}}, \ and\ \bibinfo {author} {\bibfnamefont {G.}~\bibnamefont
			{Grynberg}},\ }in\ \href@noop {} {\emph {\bibinfo {booktitle} {Atom-Photon
				Interactions: Basic Processes and Applications}}}\ (\bibinfo  {publisher}
	{Wiley},\ \bibinfo {year} {1998})\BibitemShut {NoStop}%
	\bibitem [{\citenamefont {Johnson}(1973)}]{johnson1973}%
	\BibitemOpen
	\bibfield  {author} {\bibinfo {author} {\bibfnamefont {B.}~\bibnamefont
			{Johnson}},\ }\href@noop {} {\bibfield  {journal} {\bibinfo  {journal} {J.
				Comp. Phys.}\ }\textbf {\bibinfo {volume} {13}},\ \bibinfo {pages} {445}
		(\bibinfo {year} {1973})}\BibitemShut {NoStop}%
	\bibitem [{\citenamefont {Tuvi}\ and\ \citenamefont {Band}(1993)}]{tuvi1993}%
	\BibitemOpen
	\bibfield  {author} {\bibinfo {author} {\bibfnamefont {I.}~\bibnamefont
			{Tuvi}}\ and\ \bibinfo {author} {\bibfnamefont {Y.~B.}\ \bibnamefont
			{Band}},\ }\href@noop {} {\bibfield  {journal} {\bibinfo  {journal} {J. Chem.
				Phys.}\ }\textbf {\bibinfo {volume} {99}},\ \bibinfo {pages} {9697} (\bibinfo
		{year} {1993})}\BibitemShut {NoStop}%
	\bibitem [{\citenamefont {Wang}\ and\ \citenamefont
		{Qu{\'{e}}m{\'{e}}ner}(2015)}]{wang2015}%
	\BibitemOpen
	\bibfield  {author} {\bibinfo {author} {\bibfnamefont {G.}~\bibnamefont
			{Wang}}\ and\ \bibinfo {author} {\bibfnamefont {G.}~\bibnamefont
			{Qu{\'{e}}m{\'{e}}ner}},\ }\href@noop {} {\bibfield  {journal} {\bibinfo
			{journal} {New J. Phys.}\ }\textbf {\bibinfo {volume} {17}},\ \bibinfo
		{pages} {035015} (\bibinfo {year} {2015})}\BibitemShut {NoStop}%
	\bibitem [{\citenamefont {Ye}\ \emph {et~al.}(2018{\natexlab{b}})\citenamefont
		{Ye}, \citenamefont {Guo}, \citenamefont {Gonz{\'a}lez-Mart{\'\i}nez},
		\citenamefont {Qu{\'e}m{\'e}ner},\ and\ \citenamefont {Wang}}]{ye2018}%
	\BibitemOpen
	\bibfield  {author} {\bibinfo {author} {\bibfnamefont {X.}~\bibnamefont
			{Ye}}, \bibinfo {author} {\bibfnamefont {M.}~\bibnamefont {Guo}}, \bibinfo
		{author} {\bibfnamefont {M.~L.}\ \bibnamefont {Gonz{\'a}lez-Mart{\'\i}nez}},
		\bibinfo {author} {\bibfnamefont {G.}~\bibnamefont {Qu{\'e}m{\'e}ner}}, \
		and\ \bibinfo {author} {\bibfnamefont {D.}~\bibnamefont {Wang}},\ }\href@noop
	{} {\bibfield  {journal} {\bibinfo  {journal} {Science Advances}\ }\textbf
		{\bibinfo {volume} {4}} (\bibinfo {year} {2018}{\natexlab{b}})}\BibitemShut
	{NoStop}%
	\bibitem [{\citenamefont {Gonz\'alez-Mart\'{\i}nez}\ \emph
		{et~al.}(2017)\citenamefont {Gonz\'alez-Mart\'{\i}nez}, \citenamefont
		{Bohn},\ and\ \citenamefont {Qu\'em\'ener}}]{gonzalez-martinez2017}%
	\BibitemOpen
	\bibfield  {author} {\bibinfo {author} {\bibfnamefont {M.~L.}\ \bibnamefont
			{Gonz\'alez-Mart\'{\i}nez}}, \bibinfo {author} {\bibfnamefont {J.~L.}\
			\bibnamefont {Bohn}}, \ and\ \bibinfo {author} {\bibfnamefont
			{G.}~\bibnamefont {Qu\'em\'ener}},\ }\href@noop {} {\bibfield  {journal}
		{\bibinfo  {journal} {Phys. Rev. A}\ }\textbf {\bibinfo {volume} {96}},\
		\bibinfo {pages} {032718} (\bibinfo {year} {2017})}\BibitemShut {NoStop}%
	\bibitem [{\citenamefont {Suominen}(1996)}]{suominen1996c}%
	\BibitemOpen
	\bibfield  {author} {\bibinfo {author} {\bibfnamefont {K.-A.}\ \bibnamefont
			{Suominen}},\ }\href@noop {} {\bibfield  {journal} {\bibinfo  {journal}
			{Journal of Physics B: Atomic, Molecular and Optical Physics}\ }\textbf
		{\bibinfo {volume} {29}},\ \bibinfo {pages} {5981} (\bibinfo {year}
		{1996})}\BibitemShut {NoStop}%
	\bibitem [{\citenamefont {Julienne}\ \emph {et~al.}(1994)\citenamefont
		{Julienne}, \citenamefont {Suominen},\ and\ \citenamefont
		{Band}}]{julienne1994}%
	\BibitemOpen
	\bibfield  {author} {\bibinfo {author} {\bibfnamefont {P.~S.}\ \bibnamefont
			{Julienne}}, \bibinfo {author} {\bibfnamefont {K.-A.}\ \bibnamefont
			{Suominen}}, \ and\ \bibinfo {author} {\bibfnamefont {Y.}~\bibnamefont
			{Band}},\ }\href@noop {} {\bibfield  {journal} {\bibinfo  {journal} {Phys.
				Rev. A}\ }\textbf {\bibinfo {volume} {49}},\ \bibinfo {pages} {3890}
		(\bibinfo {year} {1994})}\BibitemShut {NoStop}%
	\bibitem [{\citenamefont {Boesten}\ and\ \citenamefont
		{Verhaar}(1994)}]{boesten1994}%
	\BibitemOpen
	\bibfield  {author} {\bibinfo {author} {\bibfnamefont {H.~M. J.~M.}\
			\bibnamefont {Boesten}}\ and\ \bibinfo {author} {\bibfnamefont {B.~J.}\
			\bibnamefont {Verhaar}},\ }\href@noop {} {\bibfield  {journal} {\bibinfo
			{journal} {Phys. Rev. A}\ }\textbf {\bibinfo {volume} {49}},\ \bibinfo
		{pages} {4240} (\bibinfo {year} {1994})}\BibitemShut {NoStop}%
	\bibitem{sm}
	See Supplemental Material at the following address (to be provided by the editor)
\end{thebibliography}
\end{document}